 \documentclass[final,3p,times,twocolumn]{elsarticle}

\usepackage{amssymb,amsmath}
\usepackage{bm}

\usepackage{booktabs}
\usepackage{braket}
\usepackage{url}
\usepackage[english]{babel} 	
\usepackage{graphics} 
\usepackage{tabularx}
\usepackage{rotating}
\usepackage{float}

\graphicspath{{./figures/}}

\journal{Physics Letters B}

\begin{document}

\begin{frontmatter}



\title{Two-proton capture on the $^{68}$Se nucleus with a new self-consistent cluster model}


\author[1]{D. Hove}
\ead{dennish@phys.au.dk}
\author[2]{E. Garrido}
\author[1]{A.S. Jensen}
\author[2]{P. Sarriguren}
\author[1]{H.O.U. Fynbo}
\author[1]{D.V. Fedorov}
\author[1]{N.T. Zinner}

\address[1]{Department of Physics and Astronomy, Aarhus University, DK-8000 Aarhus C, Denmark}
\address[2]{Instituto de Estructura de la Materia, IEM-CSIC, Serrano 123, E-28042 Madrid, Spain}

\begin{abstract}
We investigate the two-proton capture reaction of the prominent rapid
proton capture waiting point nucleus, $^{68}$Se, that produces the
borromean nucleus $^{70}$Kr ($^{68}$Se$+p+p$).  We apply a recently
formulated general model where the core nucleus, $^{68}$Se, is treated
in the mean-field approximation and the three-body problem of the two
valence protons and the core is solved exactly.  The same Skyrme
interaction is used to find core-nucleon and core valence-proton
interactions.  We calculate $E2$ electromagnetic two-proton
dissociation and capture cross sections, and derive the temperature
dependent capture rates.  We vary the unknown $2^+$ resonance energy
without changing any of the structures computed self-consistently for
both core and valence particles.  We find rates increasing quickly
with temperature below $2-4$~GK after which we find rates varying by
less than a factor of two independent of $2^+$ resonance energy. The
capture mechanism is sequential through the $f_{5/2}$ proton-core
resonance, but the continuum background contributes significantly.
\end{abstract}


\end{frontmatter}

\section{Introduction}

The abundance of most stable nuclei above iron in the universe can be
understood as produced by various types of neutron
capture \cite{mat85,bar06}. However, production of about 40 stable isotopes cannot be explained in this
way, but only through similar proton capture processes
\cite{bur57,arn03,rau13,rei14,pal14}.  The basic ignition fuel is a
large proton flux arising from a stellar explosion.  The sequence of
these reactions are then one proton capture after another until the
proton dripline is reached and further captured protons are
immediately emitted. This dripline nucleus usually must wait to beta-decay to a more stable nucleus which in turn can
capture protons anew.  This is the "rp-process" \cite{sch98,bro02}.  These p-nuclei are
also believed to be produced by other methods: gamma-proton \cite{arn03,woo78} and
neutrino-proton processes \cite{fro06,fro06b}.

The beta-decay waiting time is large for some of these nuclei along
the dripline, which for that reason are denoted waiting points
\cite{oin00}. However, another path is possible to follow for
borromean proton dripline nuclei where two protons, in contrast to
one, are necessary to produce a bound nucleus.  Then two protons can
be captured before beta-decay occurs \cite{gri05,gri01}.  The capture time and the
corresponding mechanism are therefore important for the description of
the outcome of these astrophysical processes \cite{gor95,sch06}.

We focus in this letter on one of these two-proton capture reactions
leading from a prominent waiting point nucleus, $^{68}$Se
\cite{sch07,tu11}, to formation of the borromean proton dripline
nucleus, $^{70}$Kr~($^{68}$Se$+p+p$).  The specific experimental
reaction information is not available, and theoretical estimates are,
at least at the moment, unavoidable \cite{tho04,erl12,pfu12}. 

Traditionally, the reactions have been described as sequential
one-proton capture by tunneling through the combined Coulomb plus
centrifugal barrier. The tunneling capture mechanisms have been
discussed as direct, sequential and virtual sequential decay
\cite{jen10,gar05,che07,rod08}. They are all accounted for in the present formulation.
The capture rate depends on temperature through the assumed
Maxwell-Boltzmann energy distribution.  It is then important to know
the energy dependence of the capture cross section for given resonance
energies, and especially in the Gamow window \cite{chr15}.

Clearly the desired detailed description requires a
three-body model which is available and even applied to the present
processes \cite{hov16,gar04}.  However, the crucial proton-core potentials
have so-far been chosen phenomenologically to produce the essentially
unknown, but crucial, single-particle energies.  A new model
involving both core and valence degrees of freedom was recently
constructed to provide mean-field proton potentials derived from the
effective nucleon-nucleon mean-field interaction \cite{hov17,hov17b,hov17c}.  In turn
these potentials produce new and different effective three-body
potentials, which in the present letter is exploited to investigate
the two-proton capture rates. 

The techniques are in place all the way from the solution of the
coupled core and valence proton system \cite{hov16,hov17b,hov17c}, over the
self-consistent three-body input and subsequent calculations
\cite{nie01,jen04}, to the capture cross sections and rates
\cite{gar15,die10,die11,die10b}.  We shall first sketch the steps in
the procedure used in the calculations. Then we shall discuss in more
details the numerical results of interest for the astrophysical
network computations, which calculates the abundances of the isotopes in the Universe.

\section{Theoretical description}

The basic formulation and the procedure are described in
\cite{hov16,gar15,gar15b}.  The framework is the three-body technique but based
on the proton-core potential derived through the self-consistently
solved coupled core-plus-valence-protons equations \cite{hov17,hov17b}.  The procedure is
first to select the three-body method, second to formulate how to
calculate the capture rate, and finally to choose the numerical
parameters to be used in the computations.

\subsection{Wave functions} 

First, the many-body problem is solved for a mean-field treated core
interacting with two surrounding valence protons \cite{hov16}.  The
details of this recent model are very elaborate, but already applied
on two different neutron dripline nuclei \cite{hov16,hov17}.  It then
suffices to sketch the corresponding details for the present
application.  Briefly, the novel features are to find the mean-field
solution for the core-nucleons in the presence of the external field
from the two valence protons.  In turn, folding the basic
nucleon-nucleon interaction and the core wave function provides the
interaction between each valence proton and the core nucleus.  The
interactions between core and valence nucleons then depend on their
respective wave functions, which are found self-consistently by
iteration.  We emphasize that the same nucleon-nucleon interaction is
used both in the core and for this valence-proton core calculation.
The crucial main ingredient in the three-body solution is this
interaction, which then is provided by the procedure and determined
independent of subsequent applications.

The present application exploits the properties of the derived
three-body solution.  The three-body problem is solved by adiabatic
expansion of the Faddeev equations \cite{gar04} in hyperspherical coordinates. When necessary
the continuum is discretized by a large box confinement \cite{gar15a}.  The
main coordinate is the hyperradius, $\rho$, defined as the mean radial
coordinate in the three-body system \cite{gar15b,hov14}. More
specifically we have
\begin{align} 
(2m_n+m_c) \rho^2 
= m_n(\bm{r}_{v_1} - \bm{r}_{v_2})^2 
+ m_{c} \sum_{i=1}^2 (\bm{r}_{v_i} - \bm{R}_{c})^2 
\label{rhodef}
\end{align}
where $m_n$, $m_{c}$, $\bm{r}_{v_1}$, $\bm{r}_{v_2}$ and
$\bm{R}_{c}$ are neutron mass, core mass, valence-proton
coordinates and core center-of-mass coordinate, respectively.  The
three-body wave function is found through this procedure for given
angular momenta as functions of the hyperspherical coordinates for the
required ground state ($\Psi_{J}$).  When necessary the continuum is
discretized by a large box confinement and the discretized continuum
states ($\psi_{j}^{(i)}$) are calculated.

%

\subsection{Reactions}

The two-proton capture reaction $p+p+c \leftrightarrow A + \gamma$ cross section $\sigma_{ppc}$ and the photodissociation cross section $\sigma^{\lambda}_{\gamma}$ of order ${\lambda}$ are related \cite{gar15b}.
The three-body energy, $E$, and the ground state energy, $E_{gr}$,
determine the photon energy by $E_{\gamma} = E + |E_{gr}|$.  The
dissociation cross section is then given by
\begin{align}
\sigma^{\lambda}_{\gamma}(E_{\gamma}) 
=& \frac{(2 \pi)^3 (\lambda +1)}{\lambda ((2\lambda +1)!!)^2} \left( \frac{E_{\gamma}}{\hbar c}\right)^{2\lambda-1} \frac{d}{d E}\mathcal{B}({E}\lambda, 0 \rightarrow \lambda), \label{eq siggam}
\end{align}
where the strength function for the ${E}\lambda$ transition,
\begin{align}
 \frac{d}{d E} \mathcal{B}({E}\lambda, 0 \rightarrow \lambda) =
 \sum_i \left| \braket{\psi_{\lambda}^{(i)} | | \hat{\Theta}_{\lambda} | | 
 \Psi_{0}} \right|^2  \delta(E-E_i), \label{eq tran}
\end{align}
is given by the reduced matrix elements,
$\braket{\psi_{\lambda}^{(i)} | | \hat{\Theta}_{\lambda} | | \Psi_{0}}$,
where $\hat{\Theta}_{\lambda}$ is the electric multipole operator,
$\psi_{\lambda}^{(i)}$ is the wave function of energy, $E_i$, for all
bound and (discretized) three-body continuum states in the summation.
The capture reaction rate, $R_{ppc}$, is given by Ref.~\cite{die11}
\begin{align}
 R_{ppc}(E) = &  \frac{8 \pi}{(\mu_{cp} \mu_{cp,p})^{3/2}}  \frac{\hbar^3}{c^2} 
 \left( \frac{E_{\gamma}}{E} \right)^2 \sigma^{\lambda}_{\gamma}(E_{\gamma}), 
\label{eq rate}
\end{align}
where $\mu_{cp}$ and $\mu_{cp,p}$ are reduced masses of proton and
core and proton-plus-core and proton, respectively.  For the astrophysical processes in a gas of temperature, $T$, we have to average the rate in Eq.~(\ref{eq rate}) over the
corresponding Maxwell-Boltzmann distribution, $B(E,T) = \frac{1}{2}
E^2 \exp(-E/T)/T^3$,
\begin{align}
\braket{R_{ppc}(E)} = \frac{1}{2T^3} \int E^2 R_{ppc}(E) \exp(-E/T)
 \, dE, \label{eq ave rate}
\end{align}
where the temperature is in units of energy.

\subsection{Interactions}

The decisive interaction is first of all related to the mean-field
calculation of the core.  We use the Skyrme interaction SLy4 \cite{cha98}  with acceptable global average properties. The
application on one specific nuclear system requires some adjustment to
provide the known borromean character, that is unbound
proton-core $f_{5/2}$ resonance at $0.6$~MeV \cite{san14} and two protons bound to the core.  With a minimum of changes we achieve this by shifting all energies while leaving the
established structure almost completely unaltered.  The simplest
consistent such modification is by scaling all the main Skyrme
strength parameters, $t_i$, by the same factor, $0.9515$.

The density dependence of the Skyrme interaction can be viewed as a
parametrized three-body potential.  To simulate that effect we employ
a short-range Gaussian, $S_0\exp(-\rho^2/b^2)$, which depends on the three-body hyperradial coordinate $\rho$.  We choose $b=6$~fm and leave
$S_0$ to fine-tune each of the $0^+$ and $2^+$ three-body energies. This
is necessary since the keV-scale of binding is crucial for tunneling
through single MeV height barriers.  This level of accuracy is beyond
the present capability of many-body model calculations.  To reproduce
the predicted $0^+$ energy of $-1.34$~MeV \cite{wan12} a three-body
strength $S_0 = -17.5$~MeV is needed. The unknown $2^+$ energy is
varied from almost bound, zero energy, to the top of the barrier by
$S_0$ changing from $-35.05$~MeV to $-26.22$~MeV.

\section{Effective three-body potentials \label{sec:3b}}

\begin{figure}
\centering
\includegraphics[width=1\linewidth]{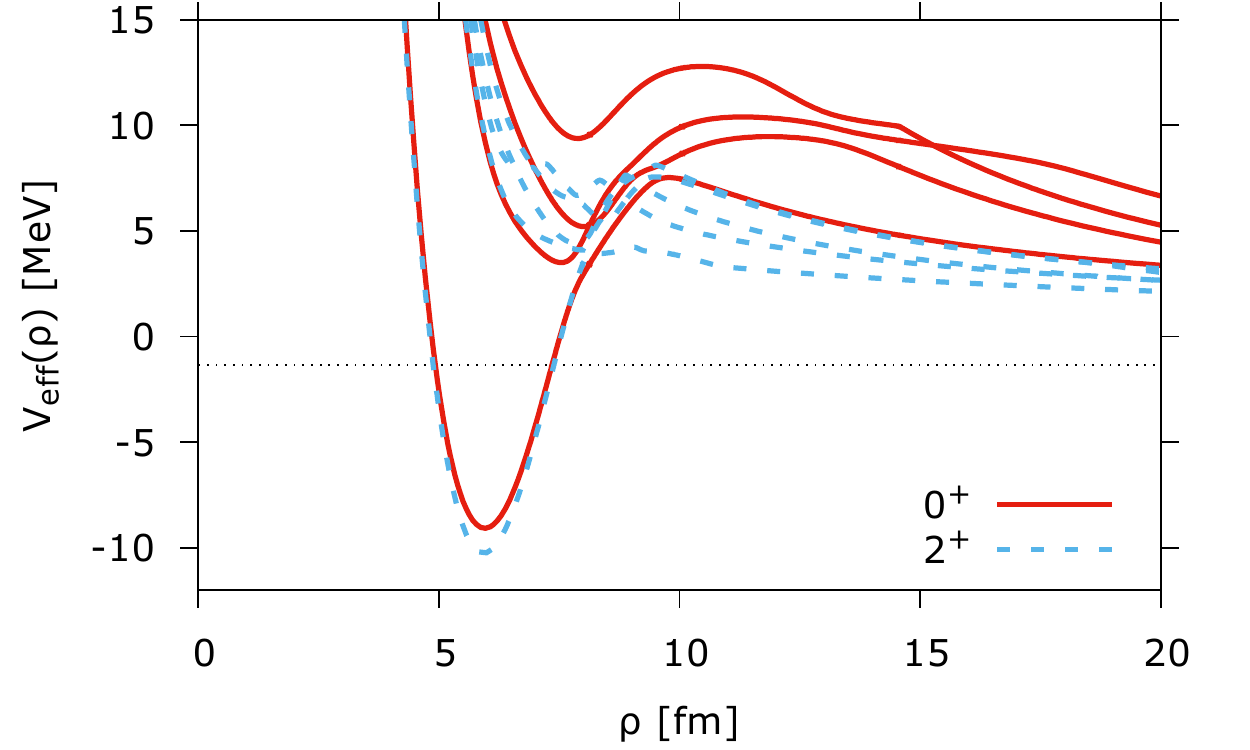}
\vspace*{-6mm}
\caption{The effective, adiabatic potentials for the $0^{+}$ (red,
  solid), and the $2^{+}$ (light-blue, dashed) configuration in
  $^{68}\text{Se}+p+p$ using the SLy4 Skyrme interaction between core
  and valence protons, scaled to reproduce the experimental $f_{5/2}$
  resonance energy of $0.6$~MeV in $^{68}\text{Se}+p$ \cite{san14}. The dotted horizontal line is the $0^+$ energy
  at $-1.34$~MeV from $S_0= -17.5$~MeV.
\label{fig:potsw}}
\end{figure}

The elaborate numerical calculations produce the sets of coupled
``one-body'' effective potentials depending on hyperradius as shown in
Fig.~\ref{fig:potsw} for both $0^{+}$ and $2^{+}$.  The continuations
beyond the $20$~fm in the figure are almost quantitatively Coulomb
plus centrifugal behavior and as such reveal no surprises.  The kinks
and fast bends reflect avoided crossings and related structure
changes.  They are especially abundant at small distances and around
the barriers.  The diagonal non-adiabatic coupling terms as well as
the diagonal structure-less three-body Gaussian potentials are included in the calculations, but not in the figure.  They both leave the structure
essentially unaltered although they may change the energies
of the solutions rather substantially.

The $0^{+}$ ground-state at $-1.34$~MeV, predicted from systematics
\cite{aud12}, is reproduced with the chosen parameters.  The structure
corresponds to the configuration of the pronounced minimum in the
lowest $0^{+}$ potential.  No  $0^{+}$ resonance are produced by the
potentials in Fig.~\ref{fig:potsw}.  The ground state is the final
state in the capture process independent of the specific mechanism.

However, the decisive capture process proceeds within the $2^{+}$
continuum from the large to the short-distance attractive region of
the potentials shown in Fig.~\ref{fig:potsw}.  This lowest minimum is
rather similar to the $0^{+}$ minimum but the non-adiabatic repulsive
terms increase the energy substantially. Unfortunately nothing is
known about a $2^{+}$ resonance which would strongly influence the
capture rate.  Consequently the strength, $S_0$, is used to vary the
position of the $2^{+}$ resonance from almost bound to disappearance
above the barriers.  Both the resonance energy, the height, and the
rather broad Coulomb shape of the barrier strongly influence the
capture process.

The structure of these potentials is substantially simpler than those
obtained in \cite{hov16} where low-lying single-proton states
$p_{3/2}$ and $f_{5/2}$ both appeared.  The present simplification is
an automatic result of the procedure using the nucleon-nucleon
mean-field effective interaction to calculate the proton-core
potential.  This is not an ad hoc assumption, but arises naturally due to
identical interactions for both core and valence particles.  As such
it is a novel deduction embedded in the design of our model.  The
lack of single-particle states of different parity implies that no
$1^-$ three-body resonance states appear in the low-energy region. The transition is then necessarily an $E2$ transition, which contributes to the longer effective lifetime of the system, and could very well be part of the reason this system is a critical waiting point.

\section{Quantitative results}

The all-important core-valence proton potential is derived naturally and unambiguously by our mean-field core treatment, as discussed in the previous sections. As a result the two-proton capture cross section follows directly, only depending on the three-body resonance level. This is discussed in the following section, after which the resulting temperature averaged reaction rates are presented. This is supplemented by a discussion of the reaction mechanism and its implication for the possible reactions.

\subsection{Cross section}

\begin{figure}
\centering
\includegraphics[width=1\linewidth]{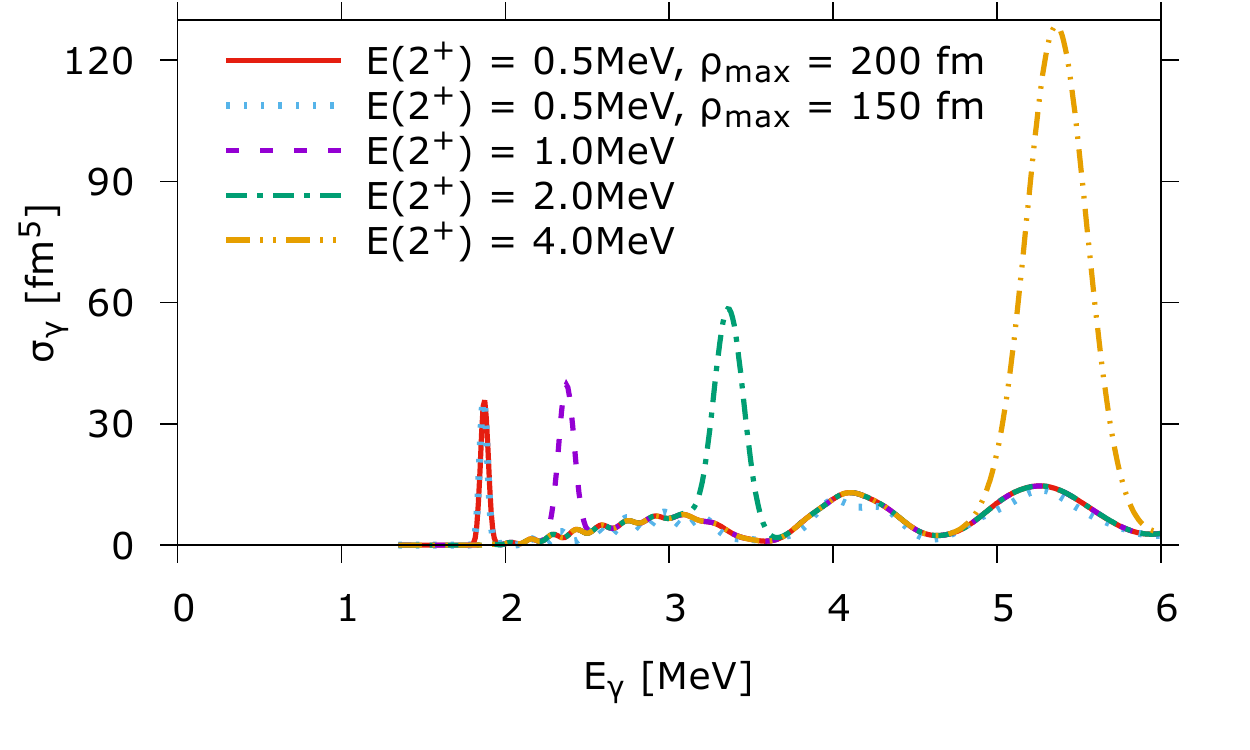}
\vspace*{-6mm}
\caption{The electromagnetic $E2$ dissociation cross section,
  $\sigma^{(\lambda=2)}_{\gamma}(E_{\gamma})$, for the proces,
  $^{70}\text{Kr} + \gamma \rightarrow ^{68}\text{Se}+p+p$, as a
  function of photon energy.  The $0^+$ final state energy is
  $-1.34$~MeV and the $2^+$ resonance energies are $E=0.5, \, 1.0, \,
  2.0,$ and $4.0$~MeV, respectively.  The discretized continuum states
  are obtained using box sizes of $\rho_{max} = 150, 200$~fm.
\label{fig:cross}}
\end{figure}

The incident flux of low-energy protons on the core nucleus may result
in capture.  The corresponding cross section is most easily obtained
from calculation of the inverse reaction, that is photodissociation of
the $0^+$ ground state, $\Psi_{0}$, of $^{70}$Kr. The discretized
continuum states, $\psi_{\lambda}^{(i)}$, are computed and the cross
section is obtained from Eqs.~(\ref{eq tran}) and (\ref{eq siggam}) with
$\lambda =2$.  The two-proton capture cross section of $^{68}$Se,
obtained from Eq.~(\ref{eq siggam}), is shown in Fig.~\ref{fig:cross} as
function of the three-body energy.

The peaks in the capture cross section occur at experimentally unknown
resonance energies where the tunneling probability is large.  We
therefore vary the energy from $0.5$~MeV to $4.0$~MeV where the widths
of the peaks in the cross section increase with energy as the top of
the barrier is approached.  We emphasize that the crucial quantity is
the resonance energy. This can be tested by varying the number of
adiabatic potentials used in the calculation. This results in
somewhat different resonance energy which however can be
compensated for by use of the three-body potential, which in turn
recover the cross section in Fig.~\ref{fig:cross}.

These features are simply understood as enhanced spatial overlaps
between the $2^+$ continuum states in the resonance region and the
ground state wave function, expressed through Eq.~(\ref{eq tran}).  Beside the resonance contributions we
also find significant, although several orders of magnitude smaller,
background contributions, which incidentally is
independent of the size of the discretization box, as long as it is sufficiently large \cite{gar15a}.

\subsection{Capture rates}

\begin{figure}
\centering
\includegraphics[width=1\linewidth]{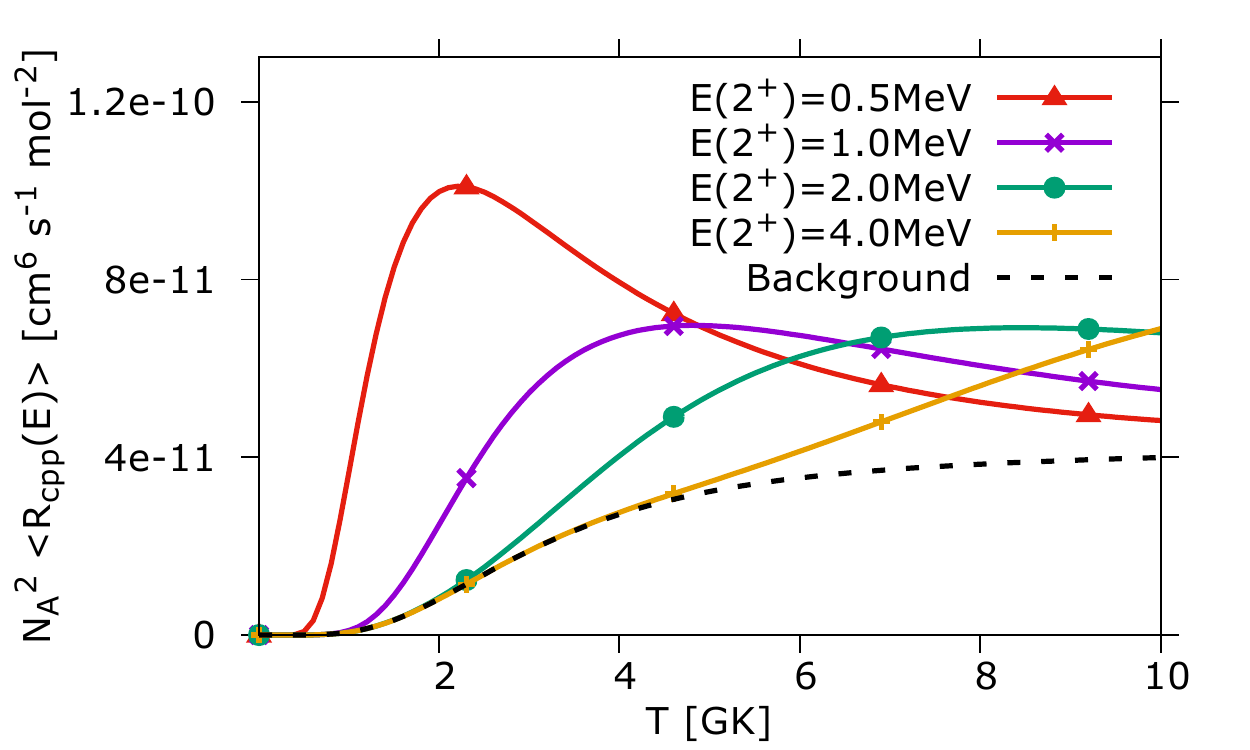}
\vspace*{-6mm}
\caption{The reaction rate for the radiative capture process
  $^{68}\text{Se}+p+p \rightarrow ^{70}\text{Kr} + \gamma$, as
  function of temperature for the different $2^+$ resonance energies
  in Fig.~\ref{fig:cross}.  The black dashed curve is the background
  contribution.
\label{fig:rate}}
\end{figure}

The capture cross sections are the main ingredient in the calculation
of the two-proton absorption rate appropriate for the temperature
dependent astrophysical network computation.  The average rate in
Eq.~(\ref{eq ave rate}) are shown in Fig.~\ref{fig:rate} as function
of temperature.  The Boltzmann smearing factor produces very smooth
curves of the same qualitative behavior.  They are zero at zero
temperature and energy, because the barrier is infinitely thick.  All
rates then increase to a maximum at the Gamow peak where the best
compromise is reached between the decreasing temperature distribution
and the increasing tunneling probability.

The peak contribution moves to higher energy and becomes smoother with
increasing resonance energy. Above temperatures of a few GK the
average rate variation is moderate and the size roughly of order
$\simeq 6 \times 10^{-11}$~cm$^{6}[N_{A} mol]^{-2} s^{-1}$.  A low-lying
resonance energy corresponds to low-lying peak position of larger
height.  We emphasize that the background without resonance
contribution obviously is smaller but only by roughly a factor of two
as soon as the temperature exceeds about $4$~GK ($\sim0.34$~MeV).  In
other words, if temperatures are in the astrophysically interesting
range below about $1$~GK, the size variations are substantial, and
vice versa above a few GK the details from the microscopic origin are
smeared out.

The actual size of the rate may reveal deceivingly little variation at
the relatively high  temperatures. However, the barrier height
and width are all-decisive and both may easily be different for other
systems where the single-particle structure at the Fermi energy is
different and perhaps more complicated as studied in \cite{hov16}.
The relatively large $2^+$ background contribution might suggest
significant corresponding $0^+$ continuum contributions.  However, the
$0^+$ barriers in Fig.~\ref{fig:potsw} are larger and the $0^+
\rightarrow 0^+$ transition as well require processes involving atomic
electrons.  It is again worth emphasizing that a superficially more
complete calculation with for example many coupled potentials would
provide the same rates after adjusting to the same resonance energy.

\begin{figure}
\centering
\includegraphics[width=1\linewidth]{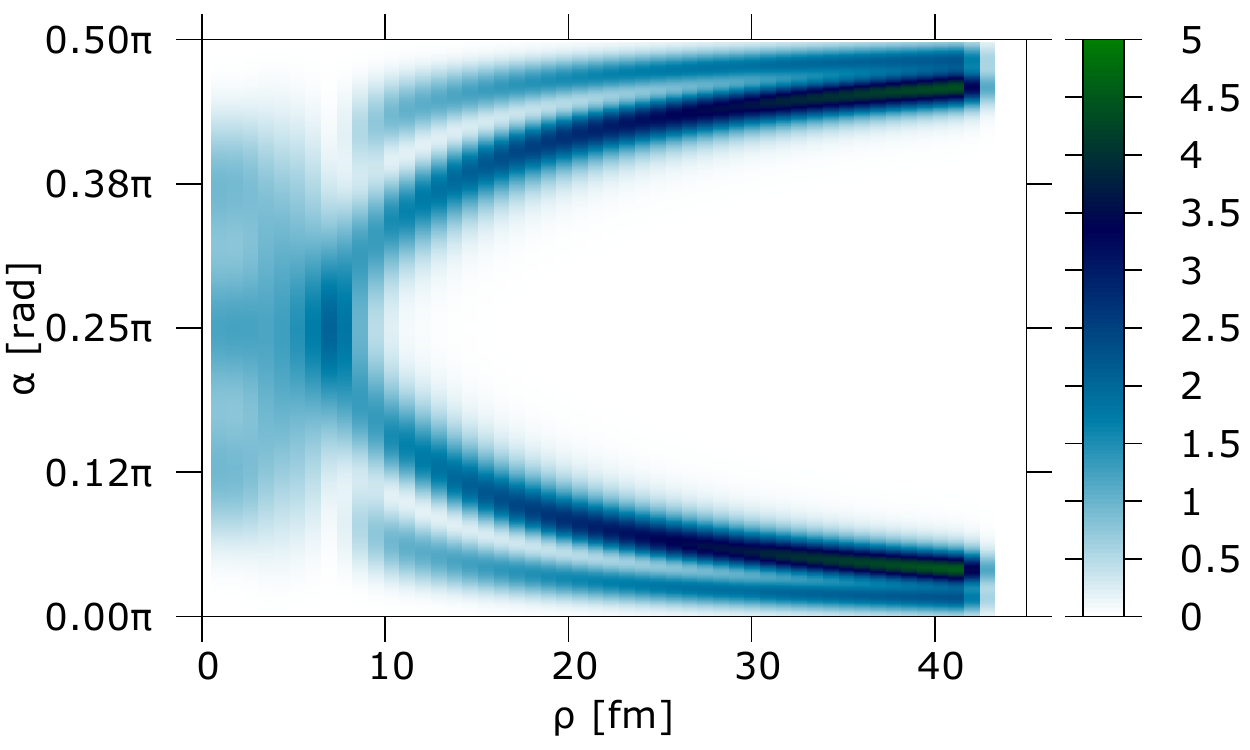}
\vspace*{-6mm}
\caption{The probability of the three-body, $^{68}\text{Se}+p+p$, wave
  function for the lowest allowed potential, integrated over
  directional angles $\left( \sin^2 (\alpha) \cos^2 (\alpha) \int
  |\Phi_n(\alpha, \rho, \Omega_x, \Omega_y)|^2 d\Omega_x
  d\Omega_y\right)$, as a function of hyperradius, $\rho$, and
  hyperangle, $\alpha$, related to the Jacobi coordinate system where
  "x" is between core and proton.  \label{fig:angDist}}
\end{figure}

\subsection{Reaction mechanism}

The rate depends on the capture mechanism.  We are here only concerned
with three-body capture, but a dense environment would enhance
four-body capture processes as discussed in \cite{die10}.  The overall
three-body process is tunneling through a barrier of particles in a
temperature distribution of given density. Once inside the relatively
thick barrier they have essentially only the option of emitting
photons to reach the bound ground-state.  However, the first of this
two-step process can occur through different mechanisms, where the
most obvious possibility is to be captured in different angular
momentum states.  The conservation of angular momentum and parity
quantum numbers are crucial in connection with resonance positions. If
low-lying $1^-$ continuum states are allowed they would be preferred,
and vice versa if prohibited $2^+$ continuum states would be
preferred.  Low-lying resonances enhance the contributions
substantially.  This selection depends strongly on the nucleus under
investigation.

For a given angular momentum of the three-body continuum states, we
still may encounter several qualitatively different ways of absorbing
two protons from the continuum \cite{gar11}. These mechanisms were discussed in
\cite{jen10} for the inverse process of dissociation, that is direct,
sequential and virtual sequential decay.  They are all accounted for
in the present formulation.  In \cite{hov16} we concluded that the
direct process is most probable for very low three-body energy when
two-body subsystems are unbound.  If the energy is larger than stable
two-body substructures such intermediate vehicles enhance the rates
and the mechanism is sequential.

Even when it is energetically forbidden to populate two-body resonance
states it may be advantageous to exploit these structures virtually
while tunneling through an also energetically forbidden barrier.  This
is appropriately named the virtual sequential two-body mechanism.  It
may be appropriate to emphasize that a similar three-body virtual
mechanism is forbidden because the three-body energy is conserved in
contrast to the energy of any two-body subsystem.

The mechanism for the present capture process is revealed in
Fig.~\ref{fig:angDist} where the $2^+$ probability integrated over the directional
angles is shown for the lowest potential as function of hyperradius
and one of the Jacobi angles.  It is a strikingly simple structure for
hyperradii larger than about $15$~fm, which for these coordinates is
equivalent to one proton at that distance from the center of mass of
the combined proton-core system.  Since the Jacobi angle, $\alpha$, is
either close to zero or $\pi/2$, this simply means that one proton is
staying very close to the core for all these hyperrradii.  Eventually
also this proton has to move away from the core since no bound state
exist. But the process is sequential through this substructure which
can be determined to be the proton-core $f_{5/2}$ resonance.

The higher-lying configurations corresponding to the three following
potentials also show precisely the same $f_{5/2}$ structure. This is
explained by combining the compact proton-core $f_{5/2}$ resonance
with one non-interacting (apart from Coulomb and centrifugal) distant
proton in any angular momentum configuration consistent with a $2^+$ structure.
The angular momenta capable of combining with $f_{5/2}$ to produce $2^+$ are $p_{1/2}, \, p_{3/2}, \, f_{5/2}, \, f_{7/2},$ and $h_{9/2}$. This also implies that for temperatures much smaller
than the $f_{5/2}$ resonance energy it would be energetically
advantageous to start the capture process in a configuration
corresponding to direct three-body capture.  The change of structure,
around avoided level crossings, to two-body resonance configurations
would then greatly reduce the barrier and substantially enhance the
capture rate.

\section{Conclusion}

The new model that treats the core and the two valence particles
self-consistently and simultaneously is applied on the waiting point
nucleus ($^{68}$Se) for the astrophysical rp-process. This is done
essentially without any free parameters or phenomenological fitting,
which makes the results much less arbitrary than usual three-body
calculations. Adding two protons, but not one, produces a bound system,
$^{70}$Kr, which is then a borromean nucleus. A moderate overall
scaling of the Skyrme interaction SLy4 reproduces the scarcely known
properties of these dripline nuclei.  Other Skyrme interactions
provide very similar results.

We calculate the radiative two-proton capture rate as function of
temperature for different resonance energies. We investigate the
mechanism and find that sequential capture of one proton after the
other by far is dominating. The first available single-particle
resonance state, $f_{5/2}$, is the vehicle, whereas the other proton
can approach in continuum states of even higher angular momentum. In practice, after tunneling through the barrier into the $2^+$ resonance
state, in practice only $E2$ electric transition to the ground state
is allowed.  Background capture through non-resonance continuum states
also contributes significantly to the capture process.  The sequential
$2^+$ capture mechanism might for other nuclei be replaced by for
example the normally larger $1^-$ capture.

In conclusion, the two-proton capture rates at a waiting point at the
dripline are successfully calculated with a conceptually relatively
simple, but technically advanced, new model.  The same effective
nucleon-nucleon interaction is used for both the nuclear mean-field
and the proton-core calculations.  The temperature dependent rate and
the corresponding capture mechanism are calculated with less ambiguity
than in previous calculations.  A number of applications are now
feasible.

\section*{Acknowledgements}
This work was funded by the Danish Council for Independent Research DFF Natural Science and the DFF Sapere Aude program. This work has been partially supported by the Spanish Ministerio de Economia y Competitividad under Project FIS2014-51971-P.


\section*{References}

\begin{thebibliography}{100}

\bibitem{mat85}  G. J. Mathews and R. A. Ward, Rep. Prog. Phys. \textbf{48}, 1371 (1985).

\bibitem{bar06} A. Barlett \textit{et al.}, Phys. Rev. C \textbf{74}, 015802 (2006). 

\bibitem{bur57} E.M. Burbridge \textit{et al.}, Rev. Mod. Phys. \textbf{29}, 547 (1957).
  
\bibitem{arn03} M. Arnould and S. Goriely, Phys. Rep. \textbf{384}, 1-84 (2003).

\bibitem{rau13} Rauscher T, Dauphas N, Dillmann I, Fröhlich C, Fülöp Z and
  Gyürky G, Rep. Prog. Phys.  \textbf{76},  066201 (2013).

\bibitem{rei14} R. Reifarth, C. Lederer, F\"{a}ppeler, J. Phys. G: Nucl. Part. Phys. \textbf{41}, 053101 (2014).

\bibitem{pal14} H. Palme, K. Lodders, A. Jones, \textit{Solar System Abundances of the Elements in Planets, Astroids, Comets and the Solar System}, A. M. Davis ed., pp 15-36 (2014).

\bibitem{sch98} H. Schatz \textit{et al.}, Phys. Rep. \textbf{294}, 167 (1998). 

\bibitem{bro02} B.A. Brown, R.R.C. Clement, H.Schatz, and A.Volya, Phys. Rev. C \textbf{65}, 045802 (2002). 

\bibitem{woo78} S.E. Woosley and W.M. Howard, Astrophys. J. Suppl. \textbf{36}, 285 (1978).

\bibitem{fro06} C. Fr{\"o}hlich, G. Mart\'{i}nez-Pinedo, M. Liebend{\"o}rfer, F.-K. Thielemann, E. Bravo, W. R. Hix, K. Langanke, and N. T. Zinner, Phys. Rev. Lett. \textbf{96}, 142502 (2006).

\bibitem{fro06b} C. Fr{\"o}hlich, P. Hauser, M. Liebend{\"o}rfer, G. Martínez-Pinedo, \textit{et al.}, Astrophys. J. \textbf{637}, 415 (2006)

\bibitem{oin00} M. Oinen et al., Phys. Rev. C \textbf{61}, 035801 (2000).

\bibitem{gri05} L.V. Grigorenko and M.V. Zhukov, Phys. Rev. \textbf{C 72}, 015803 (2005). 

\bibitem{gri01} L.V. Grigorenko, R.C. Johnson, I.G. Mukha, I.J. Thompson, and M.V. Zhukov, Phys. Rev. \textbf{C 64}, 054002 (2001). 

\bibitem{gor95} J. G{\"o}rres, M. Wiescher, and F.-K. Thielemann, Phys. Rev. \textbf{C 51}, 392 (1995). 

\bibitem{sch06} H. Schatz, Int. J. Mass spectrom. \textbf{251}, 293 (2006).

\bibitem{sch07} P. Schury \textit{et al.}, Phys. Rev. \textbf{C 75}, 055801 (2007).

\bibitem{tu11} X.L. Tu \textit{et al.}, Phys. Rev. Lett. \textbf{106}, 112501 (2011).

\bibitem{tho04} M. Thoennessen, Rep. Prog. Phys. \textbf{67}, 1187 (2004). 

\bibitem{erl12} J. Erler \textit{et al.}, Nature \textbf{486}, 509 (2012). 

\bibitem{pfu12} M. Pfützner, M. Karny, L. V. Grigorenko, and K. Riisager, Rev. Mod. Phys. \textbf{84} 567 (2012).

\bibitem{jen10} A.S. Jensen, D.V. Fedorov, and E. Garrido, J. Phys. G: Nucl. Part. Phys \textbf{37}, 064027 (2010).

\bibitem{gar05} E. Garrido, D.V. Fedorov, A.S. Jensen, and H.O.U. Fynbo, Nucl. Phys. A \textbf{748}, 27 (2005). 

\bibitem{che07} S. Chekanov \textit{et al.}, Eur. Phys. J. C \textbf{51}, 289 (2007). 


\bibitem{rod08} R. {\'A}lvarez-Rodr{\'i}guez, H.O.U. Fynbo, A.S. Jensen, and E. Garrido, Phys. Rev. Lett. \textbf{100}, 192501 (2008). 

\bibitem{chr15} C. Iliadis, \textit{Nuclear physics of stars}, John Wiley \& Sons, (2015).

\bibitem{hov16} D. Hove, A. S. Jensen, H. O. U. Fynbo, N. T. Zinner, D. V. Fedorov, and E. Garrido, Phys. Rev. C \textbf{93}, 024601 (2016).

\bibitem{gar04} E. Garrido, D.V. Fedorov, and A.S. Jensen, Phys. Rev. \textbf{C 69}, 024002 (2004). 

\bibitem{hov17} D. Hove, E. Garrido, P. Sarriguren, D. V. Fedorov, H. O. U. Fynbo, A. S. Jensen, and N. T. Zinner, Phys. Rev C  \textbf{95}, 061301(R) (2017).

\bibitem{hov17b}  D. Hove, E. Garrido, P. Sarriguren, D.V. Fedorov, H.O.U. Fynbo, A.S. Jensen, and N.T. Zinner, arXiv:1705.08718 (2017)

\bibitem{hov17c} D. Hove, E. Garrido, A. S. Jensen, P. Sarriguren, H. O. U. Fynbo, D. V. Fedorov, and N. T. Zinner, Few-Body Syst. \textbf{58}, 33 (2017).

\bibitem{nie01} E. Nielsen, D.V. Fedorov, A.S. Jensen, and E. Garrido, Phys. Rep. \textbf{347}, 373 (2001).

\bibitem{jen04} A.S. Jensen, K. Riisager, D.V. Fedorov, and E. Garrido, Rev. Mod. Phys. \textbf{76}, 215 (2004).

\bibitem{gar15} E. Garrido, Few-body Syst. \textbf{56}, 829 (2015).

\bibitem{die10} R. de Diego, D.V. Fedorov, E. Garrido, and A.S. Jensen, J. Phys. G: Nucl. Part. Phys. \textbf{37}, 115105 (2010).

\bibitem{die11} R. de Diego \textit{et al.}, Phys. Lett. B \textbf{695}, 324 (2011).

\bibitem{die10b} R. de Diego, E. Garrido, D.V. Fedorov, and A.S. Jensen, EPL. \textbf{90}, 52001 (2010).

\bibitem{gar15b} E. Garrido, A.S. Jensen, and D.V. Fedorov, Phys. Rev. C \textbf{91}, 054003 (2015).

\bibitem{gar15a} E. Garrido, Few-Body Systems \textbf{56}, 829 (2015). 

\bibitem{hov14} D. Hove \textit{et al.}, Phys. Rev. \textbf{C90}, 064311 (2014).

\bibitem{cha98} E. Chabanat, P. Bonche, P. Haensel, J. Meyer, and R. Schaeffer, Nucl. Phys. A \textbf{635}, 231 (1998).

\bibitem{san14} M.D. Santo \textit{et al.}, Phys. Lett. B \textbf{738}, 453 (2014).

\bibitem{wan12} M.Wang, G.Audi, A.H.Wapstra, F.G.Kondev, M.MacCormick, X.Xu, and B.Pfeiffer, Chin. Phys. C \textbf{36}, 1603 (2012). 

\bibitem{aud12} G. Audi \textit{et al.}, Chin. Phys. C \textbf{36}, 1287 (2012).

\bibitem{gar11} E. Garrido, R. de Diego, D.V. Fedorov, and A.S. Jensen, Eur. Phys. J. A \textbf{47}, 102 (2011). 






%
%
%
%
%
%
%
%
%
%
%
%











\end{thebibliography}
\end{document}